\documentstyle[12pt,definitions,epsfig]{article}
\begin{document}
\renewcommand{\thefootnote}{\fnsymbol{footnote}}
\pagestyle{empty}
\noindent
\begin{flushright}Imperial/TP/95-96/54 
\\ {\tt hep-th/9609040}
\\ September 1996 \end{flushright}
\vspace{3cm}
\begin{center}
{\large \bf Scheme Independence at First Order Phase Transitions }
\\ \medskip
{ \large \bf
 and the Renormalisation Group}
\\[6ex]
{\large Daniel F. Litim\footnote{E-Mail: D.Litim@ic.ac.uk}}\\[3ex]
{ Theoretical Physics Group\\ Blackett Laboratory, Prince
Consort Road\\ Imperial College, London SW7 2BZ, U.K.}\\[6ex]
\abstract{We analyse  approximate
solutions to an exact renormalisation group equation with particular
emphasis on their dependence on the regularisation scheme, which is
kept arbitrary.  Physical
quantities related to the
coarse-grained potential of scalar QED display universal
behaviour for strongly first-order phase transitions. Only subleading
corrections depend on the regularisation scheme 
and are suppressed by a sufficiently large UV scale. We calculate the
relevant coarse-graining scale and give a condition for
the applicability of Langer's theory of bubble nucleation.
}
\end{center}

\clearpage
\renewcommand{\thefootnote}{\arabic{footnote}}
\setcounter{footnote}{0}
\pagestyle{plain}
\setcounter{page}{1}

\newpage
{\bf 1.} A promising tool for the investigation of non-perturbative effects in
quantum field theory is the  Exact Renormalisation
Group (ERG) \cite{WilsonKogut74}-\cite{BHLM95}. Conceptually,  ERG
equations are quite appealing as they relate (effective) 
degrees of freedom at different length scales.  {\it Solving}
an ERG equation is  much 
more intriguing. As it couples an infinite number of operators finding
any exact solution would seem to be impossible.  Thus, any practical
computation has in addition to face the problem of finding an appropriate
truncation/approximation of the infinite dimensional space of
operators. 
However, very little is known  about the regularisation scheme (RS)
dependence of approximate solutions. 
Clearly, an approximation
scheme has to be discarded in case that a change of the RS 
either  changes the {\it qualitative} behaviour of the solution,
or introduces {\it 
large quantitative} corrections.\footnote{A recent letter
\cite{BHLM95} showed that even  critical exponents can 
depend rather strongly  on the RS.}
Therefore, it seems very useful to investigate an	 example
where the RS dependence can be made explicit. In this Letter we  study
the RS dependence of physical 
quantities related to  first order phase 
transitions at the example of scalar quantum 
electrodynamics in Euclidean space time \cite{CW73,Lawrie}. We follow the approach
advocated in \cite{Wetterich93a,ReuterWetterich93a}. 
Our conclusions agree  with   recent numerical
investigations \cite{Litim95b,Litim94a,Litim95a}. In addition, they
provide further evidence for the viability of the approximations and
yield a condition for Langer's theory of bubble nucleation
\cite{Langer} to be applicable. 

{\bf 2.} The ERG equation \cite{Wetterich93a} reads (with $t = \ln k$)
for bosonic fields  $\Phi$  
\beq \label{flow}
\0{\partial}{{\partial t}}\Ga_k[\Phi]=\012\Tr\left\{\left(\Ga^{(2)}_k[\Phi]+
R_k\right)^{-1}\0{\partial R_k}{{\partial t}}\right\} 
\eeq
where the length scale $k^{-1}$ can be interpreted as a
coarse-graining scale \cite{Wetterich93a,ReuterWetterich93a}.  
Eq.~(\ref{flow}) relates the microscopic action $S[\Phi]=\lim_{k\to
\infty}\Ga_k[\Phi]$ with 
the corresponding macroscopic (effective) action $\Ga[\Phi]=\lim_{k\to
0}\Ga_k[\Phi]$, the generating functional of 1PI Green functions. The
right hand side of \Eq{flow}  involves the regulator function $R_k$ and
the second 
functional derivative of the effective average action w.r.t.~the
fields. The trace stands for 
summation over all indices and integration over all momenta. In
momentum space, $R_k$ is a function of momentum squared $q^2$, and we
will write it, for dimensional reasons, as
\beq\label{Rk}
R_k(q^2) = q^2 g\left(\0{q^2}{k^2}\right)\ .
\eeq
In order to regularise zero modes of the second functional derivative
of $\Ga_k$, the following conditions on the RS have to be met:
\begin{center}
\begin{tabular}{lc}
(i)&$ \displaystyle \lim_{q^2\to0}\, R_k(q^2)>0$.\\[1.5ex]
(ii)&$\displaystyle  \lim_{k\to 0}\, R_k(q^2)=0$.\\[1.5ex]
(iii)&$\displaystyle \lim_{k\to \infty}\, R_k(q^2)\to \infty$.
\end{tabular}
\end{center}
Condition (i)  ensures
that $R_k$  
acts like an additional (possibly momentum dependent)  mass term  for
small momenta and
hinders  infrared divergencies in the case of massless
modes. (In the sharp cutoff limit \cite{WilsonKogut74} it reads
 $\lim_{q^2\to0}\, R_k(q^2)=\infty$.)
Condition (ii) allows to recover the usual effective action in the
limit 
$k\to 0$ as any dependence  on $R_k$ drops out. Condition (iii)
ensures that the 
classical action is obtained for  $k\to \infty$.  For any practical
applications, condition (iii) 
is weakened, replacing $k\to \infty$ by $k\to \La$ (where $\La$ is some UV
cutoff scale), and both $\La$ and $R_\La(q^2)$ have to be much 
larger than any other physical scale in the theory (typically $R_\La
\sim \La^2$).
A general class of smooth RSs is given for
arbitrary $b\ge 1$ by \cite{Wetterich93a}
\beq\label{Rksmooth}
g(y)= ({\exp y^b -1})^{-1} \ .
\eeq
Sometimes it is also very convenient to use a simple
power-like regulator \cite{derivative,Litim95b}
\beq \label{Rkmass}
g(y)= 1/y^b\ .
\eeq
It can be shown that the limit $b\to \infty$ of \Eq{Rksmooth} and
\Eq{Rkmass} corresponds to the sharp cutoff limit \cite{WilsonKogut74}. 

{\bf 3.} We now specify the flow equation for the Abelian Higgs model in $d$
dimensions (refering the reader to \cite{ReuterWetterich93a} for their
derivation), and detail the approximations used.
Our Ansatz for the relevant
part of $\Ga_k$  uses 
 the first term(s) of  a derivative expansion
\bea\label{ansatz} 
\Ga_k[\F, A]&=&\int d^dx \left\{ {U}_k(\rb)+\014  F_{\mu\nu}
F_{\mu\nu}
\right. 
 + ( D_\mu[ A]\F)^* D_\mu[ A]\F \bigg\} 
\eea
and neglects all higher derivative terms. Here, $F_{\mu\nu}$ denotes
the usual Abelian field strength, $\rb=\F^*\F$,
$D_\mu=\partial_\mu+i{\bar e} A_\mu$ the 
covariant derivative and $\bar e$ the (dimensionful) Abelian charge. 
We neglect the 
anomalous scaling of the gauge- and scalar kinetic terms and the running
of the Abelian charge, which is a valid
approximation near the Gaussian fixed point where the  anomalous
dimensions  are known to be small \cite{Litim94a,Litim95a}. 
It follows the flow
equation for the coarse-grained potential $U_k$ as
\bea\label{potflow}
&&\0{d{U}_k(\rb)}{2v_d k^{d-1}dk}=
\el d0\left(\0{{U}'_k(\rb)+2\rb\,{U}''_k(\rb)}{k^2}\right)
+
\el d0\left(\0{{U}'_k(\rb)}{k^2}\right)+(d-1) \el
d0\left(\0{2\eb2\rb}{k^2}\right)\, , 
\eea
where $v_d^{-1}=2^{d+1}\pi^{d/2}\Ga[\s0d2]$ and the primes denote the
derivatives w.r.t.~$\rb$. 
We readily identify the different terms on the r.h.s.~as the contributions from
the massive scalar/massless scalar/gauge field fluctuations
respectively. The threshold functions
\beq \label{ldn}
\el dn(w)=-
\integ{0}{\infty}{y}
y^{\s0d2+1}\0{\left(n+\de_{0,n}\right)g'(y)}{[(1+g(y))y+w]^{n+1}} 
\eeq
encode the dependence on the RS.
The integration over $y=q^2/k^2$ (momentum squared in units of $k$)
stems from the $\Tr$ on the r.h.s.~of \eq{flow}. 
Figure 1 displays the function $\el 42(\om)$ for different RS. Note
that these functions already differ in their order of magnitude for moderate
values of $\om$. 
As long as the quartic scalar self coupling
$\bar \la$ is small compared to the gauge coupling $\eb2$ we will in
addition neglect the fluctuations from the scalar field.  It
is known that 
this is the case for the Coleman-Weinberg phase transition in four
dimensions \cite{CW73,Litim94a} and for a part of the phase diagram of
normal superconductors in three dimensions where the
phase transition is first order \cite{Lawrie,Litim95b,Litim95a}.  The flow
equation then reads 
\beq\label{flow1}
\0{d {U}_k(\rb)}{dk}=2(d-1)v_d k^{d-1} \el d0\left(\0{2\eb2\rb}{k^2}\right)\, .
\eeq
This approximation can be 
controlled   comparing the gauge-field-induced
contributions to the scale dependence of $U''_k$ with those that have
been neglected. Note
also that \Eq{flow1} still allows for arbitrarily high (scalar)
couplings. No assumptions on the functional form of the potential are
made, and we do in particular
 not assume  $U_k(\rb)$ to be a local polynomial in $\rb$ (which it
is not).
However, \Eq{flow1}  will not
allow an investigation of the flattening of the inner part of the
potential in the limit $k\to 0$, which is known \cite{flattening} to
be  triggered 
by the terms $\sim \el d0(U'/k^2) + \el d0(U'/k^2+2\rb U''/k^2)$.
It follows, that the approximation becomes  invalid in the
non-convex part of the potential at some  {\it flattening scale}
$k_{\flat}>0$. We will
come back to this point later. For the time being we will study
solutions to \Eq{flow1} and their  RS
dependence with an initial potential  at some UV scale $\La$ given 
by $U_\La(\rb)= m_\La^2\rb +\s012\la_\La\rb^2 $.  
The coarse grained
potential  obtains as $U_{k,\La}(\rb)=U_\La(\rb)+
\De(\rb)$, where\beq \label{Delta}
\De(\rb)=\int_{k}^{\La}{\!\!\!\! d\bar k}\! 
\int_{0}^{\infty}{\!\!\!\!\!dy}\, 
\0{2(d-1) v_d y^{\s0d2 +1} g'(y)\, \bar k^{5}}{[g(y)+1]y\  \bar
k^2+2\eb2\rb}\, . 
\eeq
stems from integrating out fluctuations between $\La$ and $k$. 
Three different
mass scales are related to the coarse grained potential: The mass of the
scalar field in the regime with 
spontaneous symmetry breaking (SSB), $m$, the scalar mass in the symmetric
regime, $m_s$, and the mass of the gauge field in the SSB regime,
$M$. They are given by
\bea \label{masses}
m_s^2(k)&=&U'_k(\rb=0)\nonumber \\
m^2(k)&=&2 U''_k(\rb_0) \rb_0\nonumber\\
M^2(k)&=&2 \eb2 \rb_0\ , 
\eea
where $\rb_0$ denotes the location of the minimum in the SSB regime. In
the limit $k\to 0$ these masses correspond to the two-point function at
vanishing external momentum. 

{\bf 4.} To be explicit, we switch to four
dimensions, although the following discussions can be made for any
$d$.\footnote{As the gauge coupling in $d=4$ is dimensionless, we will
substitute $\eb2\to\e2$.} We will focus in the sequel on a) a
mass relation, b)  the degenerate 
(critical) potential, and  c) the surface tension.\\
{\bf a)} We start with  the  scale dependent {\it mass relation} 
\beq\label{defF}
\0{m^2}{M^2} +2 \0{\ms2}{M^2} = F\left(\0{M}{\La},
\0{k}{M},\e2\right) \, 
\eeq
with $F$ vanishing  at $k=\La$.\footnote{Any potential quadratic in 
$\bar\rho$ with $M\neq 0$ obeys  $m^2+2\ms2=0$. This relation
is best suited 
to study properties of the coarse-grained potential that are {\it
independent} of the initial conditions.}
$F$ is related to $\De(\rb)$ by
\beqx
F=\01{\e2}\left\{\De''(\rb_0)+\0{\De'(0)
-\De'(\rb_0)}{\rb_0} \right\} \ . 
\eeqx
Using \Eq{Delta} and performing the $\bar k$-integration first gives 
\beq \label{massresult}
F=
\0{3 \e2 }{8 \pi^2}\; {\cal I}_g\left[\0{2 +3 
p^2 s^2}{2[1+p^2 s^2]^2}-\left(s\leftrightarrow \01r\right) \right]\, , 
\eeq
where we have used $r=M/\La$ and $s=k/M$.
$p^2=y+yg(y)$ denotes the inverse 
effective propagator in units of $k^2$. We have also introduced the linear
integral operator 
\beq\label{Ig}
{\cal I}_g[f]=-2 \integ{0}{\infty}{y} \0{g'(y)\, f(y)}{[1+g(y)]^{3}}\, .
\eeq
that provides a RS dependent measure in momentum space.
As the argument of ${\cal I}_g$  in \Eq{massresult} depends on momenta,
it is understood that the mass relation explicitely depends on the
RS. Note, however, that 
${\cal I}_g[1]=1$ is {\it independent} of $g$, as long as $g$ fullfills the
conditions 
(i) and (ii). Furthermore, in the limit $k\to 0$ and $\La \to \infty$ 
we obtain 
\beq \label{d=4}
\0{m^2}{M^2} + 2 \0{\ms2}{M^2} = \0{3 \e2 }{8 \pi^2}
\eeq
which is 
independent of both the RS and the initial conditions.\footnote{An
analogous result was found in \cite{Lawrie}.}  
Of course, for general $k,\La$ the
r.h.s. of \Eq{massresult} is not universal and will depend on the
precise shape of $g(y)$. We have displayed the mass relation for
different RS and $\La\to\infty$ in figure 2, with $H$ given by
$3\e2 H=8\pi^2 F$. 
$H$ describes the cross-over from a "classical"  $(H\approx 0)$
region, where  fluctuations are of no importance, to a "coarse-grained"
$(H\approx 1)$  region, where fluctuations have already been
integrated out. The precise form of the cross-over 
around $k\approx M$
depends on the RS 
provided. 
Corrections due
to a finite $\La$ introduce RS dependent terms. Expanding \Eq{massresult}
for $k=0$ in powers 
of $(M/\La)^2$, we obtain
\beq
F=\0{3\e2}{8\pi^2}\sum^\infty_{n=0}(-)^n\left(1+\0{n}{2}\right) a^g_n\
r^{2n} \, .
\eeq
The leading correction is already quadratically
suppressed. The coefficients $a^g_n$ are the (RS dependent) $n^{th}$
moments of $1/p^2$ 
w.r.t.~the measure ${\cal I}_g$,
\beq \label{agn}
a^g_n={\cal I}_g\left[p^{-2n}\right]\, .
\eeq
With $g$ as in \Eq{Rkmass} they read
$a^g_n=2\Ga[1+\s0nb]\Ga[2+n-\s0nb]/\Ga[3+n]$. For large $n$, they
decay at least with $1/n$.
 A similar behaviour is observed numerically in the case of the exponential
 regulator \Eq{Rksmooth}. 
The first
subleading coefficient $a^g_1$ ranges only between  $1/3$ and
$2/3$ for  $g$ from both \Eq{Rksmooth} and
\eq{Rkmass} and with $1\le b\le\infty$. It is quite remarkable that these
coefficients are rather insensitive against a change of the RS. This is
related to the fact that only an {\it average} of the RS enters in
$a^g_n$: Assuming that $g(y)$ is monotonous, we can rewrite \Eq{agn}
as a convolution over $g$,
\beqx
a^g_n=2\int^\infty_0 dg\01{(1+g)^{3+n}}\ \01{y(g)^{n}}\ .
\eeqx
The first function in the integrand vanishes for large $g$ and has its
maximum for small $g$, whereas the second function vanishes for small
$g$ [condition (ii)] and grows large for large $g$ [condition
(iii)]. Thus, the integrand is peaked and the integral will get
contributions over 
a rather broad range of values for $g$, which
explains partly the weak RS dependence of the $a^g_n$ and is in contrast
to the RS dependence encountered in \cite{BHLM95}.\\
The
{\it coarse-graining scale} $k_\natural$ can be estimated from
\Eq{massresult}, that, 
expanded  in $s=k/M$, gives
$H=1-\s012 a^g_{-1} s^2+{\cal O}(s^4)$. For $g$ from \Eq{Rksmooth},
the coefficient $a^g_{-1}=2 \Gamma[1+\s01b]$ ranges between
$1.77$ and $2$ with $1\le b\le \infty$. The plateau is reached at the
1\% level  as soon as the 
coarse-graining scale is about 10\% of the gauge-field mass,
\beq \label{coarse}
k_\natural  \approx 0.1\, M\ .
\eeq \\
{\bf b)} The {\it critical potential}
$U^{\rm crit}_{k,\La}$ corresponds to the one with degenerate minima,
$U^{\rm crit}_{k,\La}(0)$ = $U^{\rm crit}_{k,\La}(\rb_0)$. 
The  initial conditions, that lead to a degenerate
potential with v.e.v.~$\rb_0$ at some scale $k$ are uniquely specified
as
\bea\nonumber
m^2_{\La,\rm c} &=&
\Delta'(\rb_0)-\02{\rb_0}\left\{\Delta(\rb_0)-\Delta(0)\right\}\\
\la_{\La,\rm c} &=&-\02{\rb_0^2}\left\{\rb_0
\Delta'(\rb_0)-\Delta(\rb_0)+\Delta(0)\right\}\, , \label{crit}
\eea
and we obtain
\bea	
U^{\rm crit}_{k,\La}(\rb)&=&
\0{\rb}{\rb_0}[\Delta(\rb_0)-\Delta(0)]\left(\0{\rb}{\rb_0}-2\right)
+\Delta'(\rb_0)\rb\left(1-\0{\rb}{\rb_0}\right)
+\Delta(\rb)\ .
\eea
Using \Eq{Delta} and \Eq{Ig}, the critical potential follows
(apart from an irrelevant field-independent constant) as
\bea
U^{\rm crit}_{k,\La}(\rb)&=&6 v_4 e^4 \left\{ {\cal I}_g\left[
\rb^2\, \ln \0{2\e2 \rb +p^2 k^2 }{M^2+p^2 k^2 }\right. \right.
+
\left.\left.\0{M^2\rb(\rb_0-\rb)}{M^2+p^2k^2} \right] -
(k\leftrightarrow\La)\right\}. \label{Ucrit}
\eea
Again in the limit $k\to 0, \La\to\infty$ any RS
dependence of the critical potential drops out, and, using $z=
\rb/\rb_0$, we are left with the potential
\beq \label{Uuniversal}
U^{\rm crit}_{0,\infty}(\rb)=\0{3 M^4}{64 \pi^2} \left\{
z^2\ln z
+z \left(1-z\right)\right\} \ 
\eeq
and $m^2/M^2=6 v_4 \e2$. 
Corrections due to a finite  $\La$ can be expanded as a power
series in $(M/\La)^2$. For $k=0$ we obtain
\bea \label{UcritLa}
&&U^{\rm crit}_{0,\La}=U^{\rm crit}_{0,\infty}-\032 v_4 M^4 
\sum^{\infty}_{n=0}a^g_{n+1}\ p_n(z)z(1-z)^2 \0{(-)^n r^{2n+2}}{n+\de_{n,0}}
\eea
where $p_n$ are $n^{th}$-order polynomials in $\rb/\rb_0$,
\beqx
p_n(z)=\sum_{m=0}^{n}(m+1)z^{n-m}\ .
\eeqx
{\bf c)} The {\it surface tension}
$\si$, defined as  
\beq\label{sigma}
\si_{k,\La}(\rb_0) = \int^{\bar \F_0}_0\!\!\! d \F \sqrt{2 U^{\rm
crit}_{k,\La}(\rb)}  
\eeq
is sensitive to the shape of the critical potential. 
We expect on dimensional grounds that 
it scales  like  $\si \sim \rb_0^{\ 3/2}$, and with
\Eq{Uuniversal} we obtain 
\beq\label{sigma00}
{\si}={\sqrt{3 v_4}\,a_0\e2\,\rb_0^{\ 3/2}}= \0{3\,a_0}{32
\pi^2}\0{M^4}{m} \ . 
\eeq
The coefficient $a_0=\int^1_0dx\sqrt{1-x+x\ln x}\approx 0.42$ encodes the
shape of the critical potential. 
Defining $\de\si=\si-\si_{0,\La}$ we again find that RS
dependent terms are suppressed by powers in 
$(M/\La)^2$,
\beq
\0{\de\si}{\si}=\sum^\infty_{n=1}(-)^{n+1}\ a^g_n\,\0{a_n}{a_0}\,
r^{2n} \ . \eeq
The expansion coefficients depend on the shape of the critical potential
encoded in the 
numerical factors $a_n$ with $a_1=\int_0^1dx(1-x)^2/\sqrt{x\ \ln
x+1-x}\approx 0.30$ (and similar expressions for the higher terms
$a_2\approx 0.24, 
a_3\approx 0.18,\ldots$), and on the 
scheme dependent numbers $a^g_n$. In figure 3 we have displayed both
the v.e.v.~$\rb_0$ and the surface 
tension as functions of $m/M$. 

{\bf 5.} The small-momentum fluctuations of the scalar field become
important for the non-convex part of the potential. From \Eq{potflow} we
can estimate this scale through $k_{\flat}^2\approx {\rm
max}\{-U',-U'-2\rb U''\}$ and obtain with \Eq{Uuniversal}
\beq \label{flat} k_{\flat}^2 \approx 3 v_4  \e2 M^2 \ .\eeq
Evaluating the mass relation \Eq{massresult} at $k=k_{\flat}$ instead of $k=0$
induces subleading corrections $\sim v_4 \e2$ to \Eq{d=4}. This is consistent
with an estimation using  the
partial differential equation \eq{potflow}.     
Defining $\rho=k^{2-d}\bar\rho$, $u(\rho)=k^{-d} U(\bar\rho)$,
$\e2=k^{d-4}\eb2$ and 
expanding the r.h.s.~of the flow equation for $u(\r)$ up to
linear order in $u'(\r)$, we obtain 
\bea 
 \0{\partial u}{\partial t}&=& -d\, u +\left[(d-2)\rho-4v_d\el d1(0)\right]
\0{\partial u}{\partial \r}
+4v_d\el d0(0)+ 2(d-1)v_d \el d0(2\e2\rho)
\, . \label{pde}
\eea
The corresponding mass
relation \Eq{massresult} obtains a more 
complicated $\e2$-dependence that reads in the sharp cutoff limit
$m^2/M^2 +2 m_s^2/M^2 = 3\e2/8\pi^2/(1+2 v_4 \e2)^2$
\cite{Litim95b,Litim94b}.  

It remains to be shown that the surface tension  has already reached a
plateau at the coarse-graining scale $k_{\natural}>k_\flat$
before the scalar fluctuations are relevant.
This is an important prerequisite for Langer's theory of
bubble nucleation \cite{Langer} to be applicable, and of relevance for
the electroweak phase transition \cite{EW,EWnick}. 
In figure 4 the surface tension is calculated as a function of
$k/M$ for fixed v.e.v.~$\rb_0$ and different RSs. The initial 
conditions 
are chosen as 
to give a degenerate potential with v.e.v.~$\rb_0$ at some scale $k$
with the corresponding surface tension $\si_k$. In the limit
$k\to 0$ the critical initial values \Eq{crit} are independent of
$k$. Decreasing $k$ shows
that $\si$ initially increases until it reaches a constant value.  The  ratio
of flattening to 
coarse-graining scale obtains from \Eq{coarse} and \Eq{flat} as
\beq		\label{ratio}		
k^2_\flat/k^2_\natural \approx \e2
\eeq
and is effectively RS independent.\footnote{The uncertainty in defining
the scales $k_\flat$ and $k_\natural$ is much larger than the RS dependence.}
The condition for a coarse-grained surface tension to be a
well-defined quantity ($k_\flat/k_\natural<1$) is automatically
fullfilled with $e^2\ll 1$. Ultimately, this is related
to the strength of the phase transition, being strongly first
order. 
For weakly first order transitions we have to expect that
$k_\flat/k_\natural \sim \C{O}(1)$ or even larger, and  defining
a coarse-grained surface tension becomes ambiguous.\footnote{See,
however, the discussion for 3d matrix models in \cite{Berges}, where
also a result analogous to our \Eq{ratio3} is given.}

Our results are by no means specific to the four dimensional 
case. The main difference in three dimensions 
comes from the fact that the gauge coupling has the dimension of a
mass. Furthermore, the scale dependence of $\eb2$ is in general no
longer negligible and a non-trivial fixed 
point structure governs the flow equation
\cite{Litim95b,Litim95a,Litim94c}. Our approximations  remain valid as
long as the gauge coupling stays 
in the vicinity of the Gaussian fixed point. The mass relation
\Eq{d=4} then becomes 
\beq\label{3dmass}
m^2+2 m_s^2 =\01{2\pi} \eb2 M\, 
\eeq
and the critical potential \Eq{Uuniversal} reads
\beq
U=\0{M^3}{12 \pi} \left[ z(1+z) - 2z^{3/2}\right]\ .
\eeq
Note the appearence of the non-analytic term $\sim |\F|^3$ in the
potential. The condition \Eq{ratio} generalises to
\beq \label{ratio3}
k^2_\flat/k^2_\natural \approx \0{\eb2}{M}\ .
\eeq  
With $\eb2/M$ being the perturbative expansion parameter it follows
that Langer's theory is viable within the perturbative regime.
 A full discussion where the running gauge coupling is also taken into
account  will be given elsewhere \cite{Litim96b}.  

{\bf 6.} In summary, we calculated physical quantities at a photon
induced first
order phase transition and  studied their RS (in-)dependence.  The main RS
dependence enters through the 
finite  UV scale $\La< \infty$  and the scale $k_\flat>0$, both of
which introduce terms proportional to powers of $k_\flat/M$ or $M/\La$, with
$M$ being a typical mass scale of the theory.  The related
expansion coefficients $a^g_n$ show a very weak RS dependence which
indicates the stability of the approximation used. Especially, as the
RS dependent 
terms introduce only marginal quantitative corrections, none of  the  physical
conclusions are affected by them. This is of immediate relevance for
the sophisticated numerical investigations presented in
\cite{EWnick,Berges}. Note also
that these RS dependent terms would not show up at a second order phase
transition, where a scaling ({\it i.e.}~$k$-independent) solution is
obtained. 
It emerged that the concept of a
coarse-grained surface tension based on the seperation of low- and
high energy modes is only viable for sufficiently
strongly first order phase transitions. The criterion for the
validity of the standard treatment of bubble nucleation is RS
independent. \\[1ex] 
 
We wish to thank 
C.~Wetterich for discussions. This work was
supported in part by the European Commission under the Human Capital
and Mobility programme, contract number CHRX-CT94-0423.

\newpage
\begin{figure}[t]
\begin{center}
\unitlength=0.0014\epsfxsize
\begin{picture}(700,500)
\put(710,20){\huge $\om$}
\put(130,90){\fbox{\huge  $\el 42(\om)$}}
\put(390,400){\scriptsize
\begin{tabular}{cc}
Sharp Cutoff &$\thicklines  \put(0,0){\line(44,0){44}}${}\\
$g=y^{-1}${}&$\thicklines  \multiput(0,0)(16,0){3}{\line(12,0){12}}$\\
$[\exp y-1]^{-1}$&$\thicklines  \multiput(12,0)(18,0){2}{\line(2,0){2}}
\multiput(0,0)(18,0){3}{\line(8,0){8}}${}\\
$y^{-2}$&$\thicklines \multiput(1,0)(4,0){11}{\line(2,0){2}}${}
\end{tabular}}
\epsffile{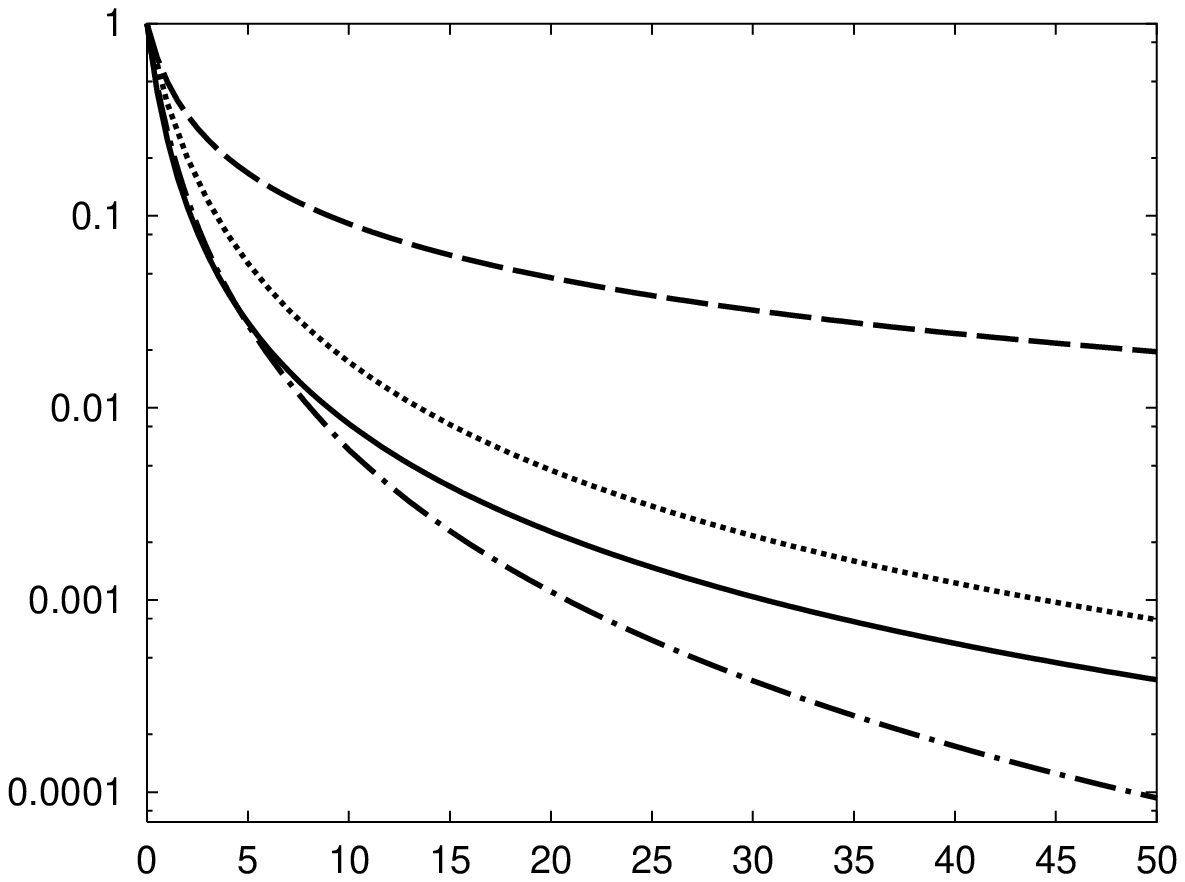}
\end{picture}
\end{center}
\caption{The threshold function $\ell^4_2(\omega)$ for
different RS.}
\end{figure}
\begin{figure}[t]
\begin{center}
\unitlength=0.0014\epsfxsize
\begin{picture}(700,500)
\put(710,20){\huge $\frac{k}{M}$}
\put(140,150){\fbox{\huge  $H\!\left(\frac{k}{M}\right)$}}
\put(390,400){\scriptsize
\begin{tabular}{cc}
Sharp Cutoff&$\thicklines  \put(0,0){\line(44,0){44}}${}\\
$g=y^{-1}${}&$\thicklines  \multiput(0,0)(16,0){3}{\line(12,0){12}}$\\
$[\exp y-1]^{-1}$&$\thicklines  \multiput(12,0)(18,0){2}{\line(2,0){2}}
\multiput(0,0)(18,0){3}{\line(8,0){8}}${}\\
$y^{-2}$&$\thicklines \multiput(1,0)(4,0){11}{\line(2,0){2}}${}
\end{tabular}}
\epsffile{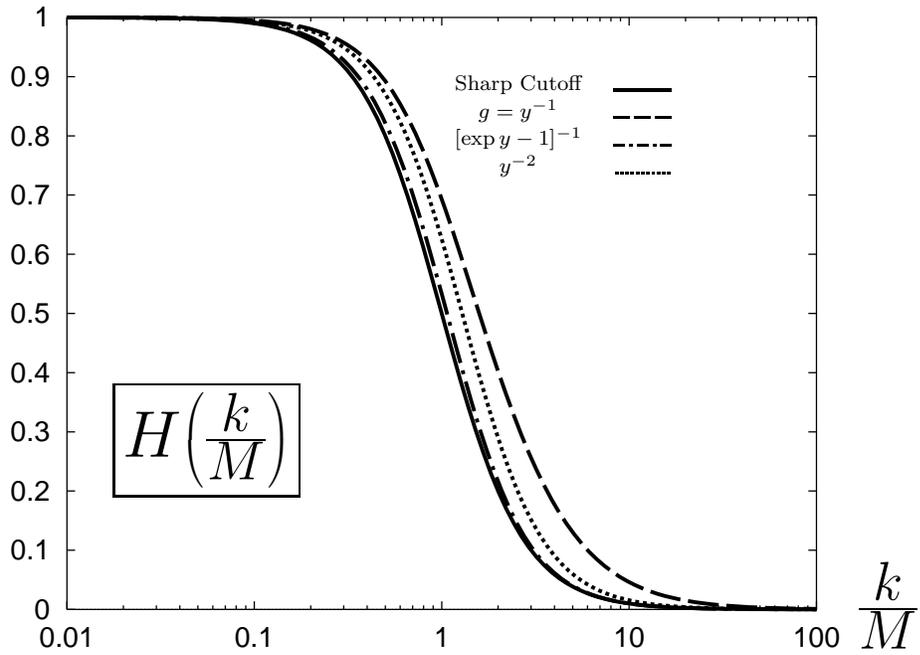}
\end{picture}
\end{center}
\caption{The mass relation for different RS.}
\end{figure}
\begin{figure}[t]
\begin{center}
\unitlength=0.0014\epsfxsize
\begin{picture}(700,500)
\put(710,20){\huge $\frac{m}{M}$}
\put(430,370){\large
\begin{tabular}{cc}
$\displaystyle \frac{\bar\rho_0}{M^2}$&$\thicklines
\put(0,0){\line(44,0){44}}${}\\[2ex] 
$\displaystyle \frac{\sigma}{M^3}$&$\thicklines
\multiput(0,0)(16,0){3}{\line(12,0){12}}$ 
\end{tabular}}
\epsffile{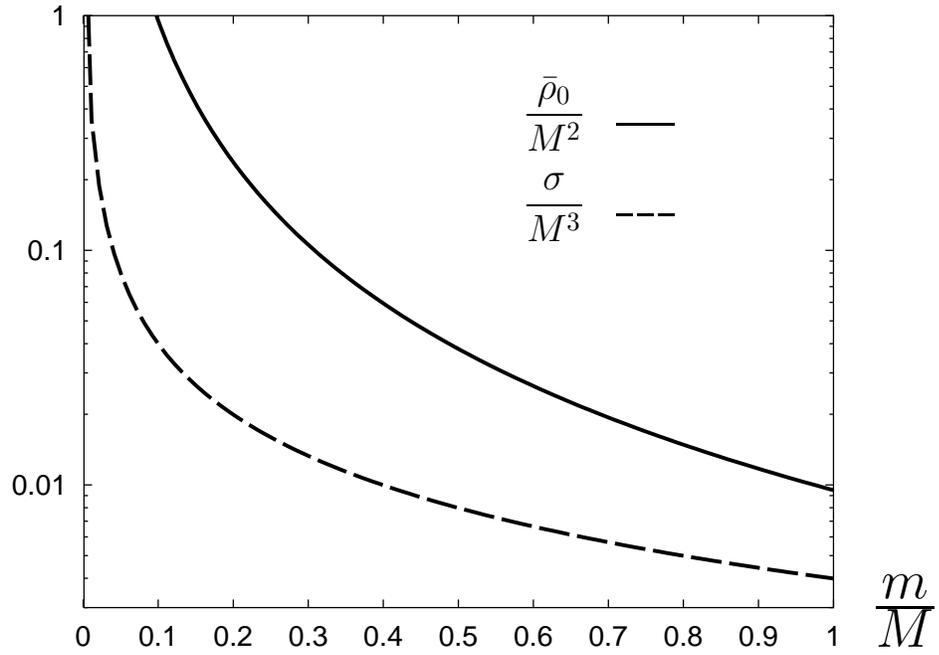}
\end{picture}
\end{center}
\caption{The surface tension $\si$ and the v.e.v.~$\bar\rho_0$ as functions
of $m/M$.}
\end{figure}
\begin{figure}[t]
\begin{center}
\unitlength=0.0014\epsfxsize
\begin{picture}(800,500)
\put(710,20){\huge $\frac{k}{M}$}
\put(150,150){\fbox{\Huge  $\frac{\sigma_k}{\sigma_0}$}}
\put(390,400){\footnotesize
\begin{tabular}{cc}
Sharp Cutoff&$\thicklines  \put(0,0){\line(44,0){44}}${}\\
$g = y^{-1}${}&$\thicklines  \multiput(0,0)(16,0){3}{\line(12,0){12}}$
\end{tabular}}
\epsffile{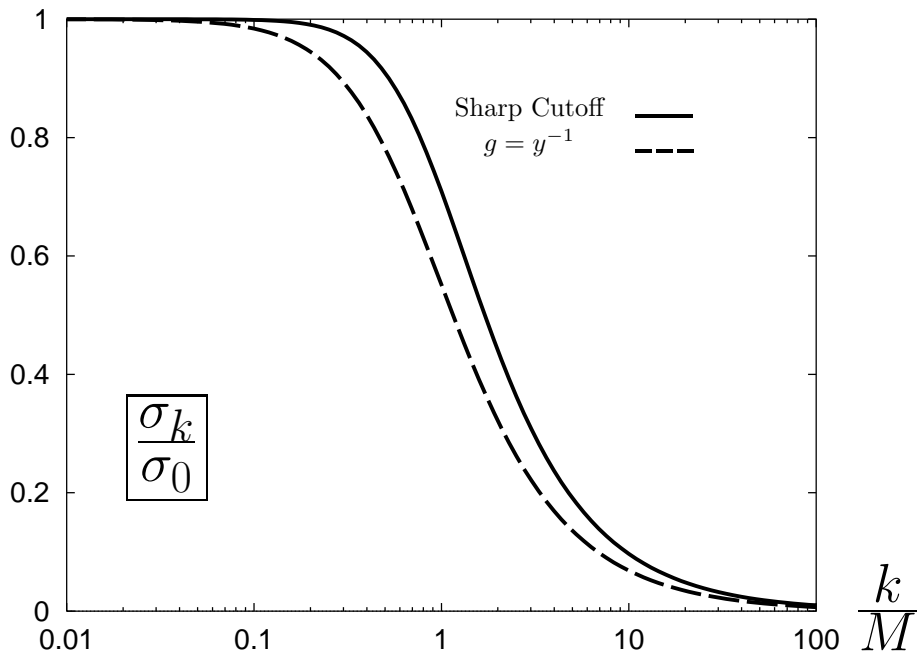}
\end{picture}
\begin{minipage}{.8\hsize}{\small
\caption{The surface tension for fixed v.e.v.~$\bar\rho_0$ as a
function of $k/M$ for different RS in the limit $\La\to\infty$.}}
\end{minipage}
\end{center}
\end{figure}

\end{document}